\documentstyle[twocolumn,prb,aps,epsf]{revtex}
\begin{document}
\draft
\title{ Shot Noise Enhancement in Resonant Tunneling Structures in a Magnetic
Field}
\author{V. V. Kuznetsov, E. E. Mendez}
\address{
Department of Physics and Astronomy, State University of New York at Stony
Brook, Stony Brook, NY 11794-3800}
\author{J. D. Bruno and J. T. Pham}
\address{
Army Research Laboratory, Adelphi, MD 20783}
\date{\today}
\maketitle

\begin{abstract}
We have observed that the shot noise of tunnel current, $I$, in
GaSb-AlSb-InAs-AlSb-GaSb double-barrier structure under a magnetic field can
exceed $2qI$. The measurements were done at T=4K in fields up to 5T parallel to
the current. The noise enhancement occurred at each of the several
negative-differential conductance regions induced by the tunneling of holes
through Landau levels in the InAs quantum well. The amount of the enhancement
increased with the strength of the negative conductance and reached values  up
to $8qI$. These results are explained qualitatively by fluctuations of the
density of states in the well, but point out the need for a detailed theory of
shot noise enhancement in resonant tunneling devices.
\end{abstract}


\narrowtext

Since shot noise is sensitive to correlation effects that result from Pauli
exclusion principle and the Coulomb interaction between charged particles, noise
measurements can give information about the kinetic of electrons in a conductor.
The realization of this fact has spurred recent interest in shot noise in a
variety of systems \cite{review}, including the double-barrier
resonant-tunneling diode (RTD).

The current-voltage characteristic (I-V) of an RTD usually has a
quasi-triangular shape, with an initial region of positive differential
conductance (PDC) before the current peak, followed by a sharp region of
negative differential conductance (NDC) just after the peak.  So far, most
studies of shot-noise have focused on the PDC region, probably because
experimentally the NDC region is often masked by an external instability that
depends on the circuit to which the diode is connected.  Measurements in
GaAlAs-GaAs-GaAlAs RTDs have shown a significant noise suppression relative to
the "full" shot noise $2qI$ ($q$ is the electron charge and I the tunneling
current), which has been explained by correlation between tunneling electrons
\cite{Li}.

The deviation of the actual shot noise from $2qI$ is quantified by a Fano
factor, defined as the ratio of the noise spectral density, $S_I(\omega)$, to
the full shot noise.  Using either a quantum-transport \cite{theory1} or a
semiclassical theory \cite{theory2} it has been shown that the Fano factor can
be expressed in terms of the tunneling probabilities $T_1$ and $T_2$, through
the two potential barriers:
\begin{equation}
\gamma\equiv\frac{S_I(\omega)}{2qI}=1-\frac{2T_1T_2}{(T_1+T_2)^2}.
\label{eq1}
\end{equation}
The value for $\gamma$ ranges from $\gamma=1$, for very asymmetric barriers, to
$\gamma=1/2$, when $T_1=T_2$. For completely incoherent transport we can view
the two barriers as two resistances in series and if each resistance generates
full shot noise then tunneling probabilities in (1) are replaced by differential
resistances of the barriers \cite{Beenakker}. As Landauer  \cite{Landauer} has
pointed out, the maximum noise suppression is readily obtained then for two
identical diodes in series.

The few results available in the NDC region exhibit a behavior very different
from the suppression found in the PDC region.  Li et al.\cite{Li} showed
enhancement of noise in the NDC region, although, their focus being on noise
suppression in the PDC region, they did not discuss that result in detail.
Brown \cite{Brown} has reported an enhancement of the current noise for a single
voltage in the NDC region, confirmed by Iannaccone et al. \cite{Iann98} in
measurements for voltages throughout the entire NDC region. They have explained
their observation in terms of correlation produced by shifts in the density of
states for tunneling, and have accounted for the noise enhancement with a
calculation that extends a formalism originally developed for noise suppression
in the PDC region \cite{Iann97}.Theoretically, Pytte and Thomas studied long ago
the fluctuations in Gunn diodes, whose I-V characteristics are similar to those
of resonant-tunneling structures, but they did not consider in detail the NDC
region \cite{old}.

In this paper we report the observation of multiple peaks in the noise
characteristics of a RTD when a magnetic field is applied parallel to the tunnel
current.  Similarly to previous works, we find that whenever there is an NDC
region in the I-V characteristics of the diode the noise is enhanced over its
background and shows a peak.  But the fact that in our experiments the number of
NDC regions and their strengths depend on the intensity of the magnetic field
has allowed us to correlate directly the enhancement of shot noise with the
corresponding NDC values, which should be valuable to theories aiming at
explaining in detail the enhancement of shot noise.

The experiments have been done in a type II GaSb-AlSb-InAs-AlSb-GaSb
double-barrier structure. The top of the valence band of GaSb (electrode) lies
above the bottom of the conduction band of InAs (quantum well).  As a
consequence, there is a charge transfer between the two materials, with
two-dimensional (2D) electrons accumulating in the InAs well and comparable
number of 2D holes accumulating on the GaSb side of the GaSb-AlSb interfaces.
AlSb barriers provide charge confinement and separation.  When a voltage is
applied between the GaSb electrodes, holes tunnel from the  emitter to the
collector via a quantum state in the conduction band of InAs \cite{notes} (See
upper inset of Fig. 1).  With increasing voltage more states are available for
tunneling and the current increases, until misalignment between states prevents
further tunneling and the current abruptly goes to zero \cite{Mendez}. Our
structure is different from the conventional GaAs/AlGaAs heterostructure in that
at large bias the quasi-Fermi level in the well is always above the valence band
of the emitter. Thus, it is possible for holes tunneling into the well to be
scattered inelastically above the emitter valence band, in which case Pauli
exclusion principle has no effect and correlation is only due to Coulomb
repulsion.

\begin{figure}
\epsfclipon
\epsfxsize 0.5\textwidth
\epsffile{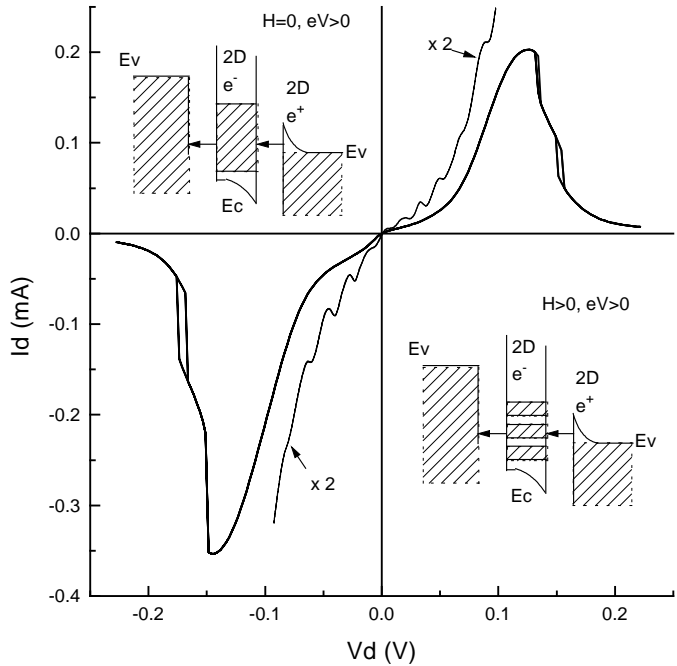}
\vspace{0.2cm}
\caption{Experimental I-V curve of GaSb/AlSb/InAs RTD at H=0 and H=3T (line,
multiplied by 2 for clarity). Top insert: schematic band profile of the
heterostructure under positive bias, at H=0. The arrows show tunneling holes.
Bottom insert: the same structure when a magnetic field is applied. The density
of states in the well is modified by a periodic Landau level structure.}
\end{figure}

Figure 1 shows the experimental I-V characteristics, at T = 4.2K, for a 20$\mu$m
diode that consists of the following regions: 3,000\AA\ p$^+$ GaSb
(N$_a$=2x10$^{18}$ cm$^{-3}$); 500\AA\ p GaSb (1x10$^{17}$ cm$^{-3}$); 50\AA\
undoped GaSb; 50\AA\ undoped AlSb; 150\AA\ undoped InAs; 50\AA\ undoped AlSb;
50\AA\ undoped GaSb; 500\AA\ p GaSb (1x10$^{17}$ cm$^{-3}$), (2x10$^{18}$
cm$^{-3}$)  p$^+$ GaSb substrate.  The small current asymmetry between both
voltage polarities in Fig.1 is possibly due to an effective asymmetry between
the two AlSb barriers.  The hysteresis regions are from instabilities in the
circuit external to the diode and will not be discussed here.

Also shown in Fig.1 is the I-V characteristic when the diode is subject to a
magnetic field parallel to the tunnel current.  At the Landau-level energies the
field creates sharp maxima in the density of states of the well and deep minima
(or gaps) in between (See lower inset of Fig. 1). As a result, the I-V
characteristic exhibits distinct features associated with tunneling via
individual Landau levels. These features become more pronounced with increasing
field and eventually show NDC, as illustrated in Fig. 1 for H = 3T. Moreover,
for a fixed field, especially if it is weak or moderate, the strength of those
features varies. This variation (whose origin is not clear), combined with that
offered by the change in field, allows for a wide range of stable NDC regions.

The shot noise of the diode was measured at T = 4K, with the device grounded at
one end and in series with a load resistor R$_0$ close to it.  (The value of
R$_0$ was 420 $\Omega$ up to a field of 3T, and 77 $\Omega$ for 4T and 5T.) The
sample was connected with low-impedance coaxial cables to a room-temperature
circuit which included a R$_1$ = 110 k$\Omega$ resistor through which an AC
signal was applied to the sample and measured simultaneously with the noise
spectrum. The voltage noise across the device in parallel with the load resistor
was measured by a low voltage noise preamplifier (voltage noise of
1.4nV/Hz$^{1/2}$ and noise resistance of about 1k$\Omega$) followed by a
spectrum analyzer with a 100kHz frequency band.

The procedure for calculating the current noise of the device was as follows.
First the voltage noise of the preamplifier and the leads (measured separately)
was subtracted from the output noise, and then the result was divided by the
normalized AC signal, which is the product of the differential resistance and
coefficient of amplification. Finally, the small background current noise of the
preamplifier was subtracted. The spectrum of the noise was "white" up to 100kHz.
When repeated for both directions of the voltage sweep (each took several
hours), the experimental curves were reproducible, which indicates stability of
both the sample and the electronics.

At zero magnetic field and small tunnel currents, the measured shot noise for
the PDC region follows the value for uncorrelated electrons ($2qI$), as shown in
Fig. 2.  For larger currents the noise is significantly suppressed, but
eventually increases, crosses the classical value and then is enhanced over the
$2qI$ value.  For positive bias the suppression is about one-half of the full
shot noise, as expected for a symmetric double barrier structure \cite{Liu}, but
for negative bias the suppression is larger. Although we do not know the origin
of  this large suppression, we have no reason to believe that is related to the
materials configuration of our heterostructure. To our knowledge, such a large
suppression has been reported only once \cite{Brown} and, with the exception of
a numerical calculation \cite{Jahan}, existing theories do not predict noise
reduction larger than one-half \cite{Iann97,Chen}.

In the presence of a small field the noise in the diode is modified in two ways,
as seen in Fig.2 for H = 3T. 
There is a deviation of the 3T background from H=0
curve, especially at low currents, where the field suppresses shot noise beyond
$2qI$. The most visible change is the appearance of sharp peaks superimposed on
a noise background. Some of the noise peaks are as large as to even exceed the
full shot noise value.  The pronounced loop of the peak at -0.05mA indicates
that the current itself peaks at this value in the I-V characteristic, as is
better seen on Fig. 3, where the measured noise is plotted as a function of bias
along with the quantity $2eI$.

\begin{figure}
\epsfclipon
\epsfxsize 0.5\textwidth
\epsffile{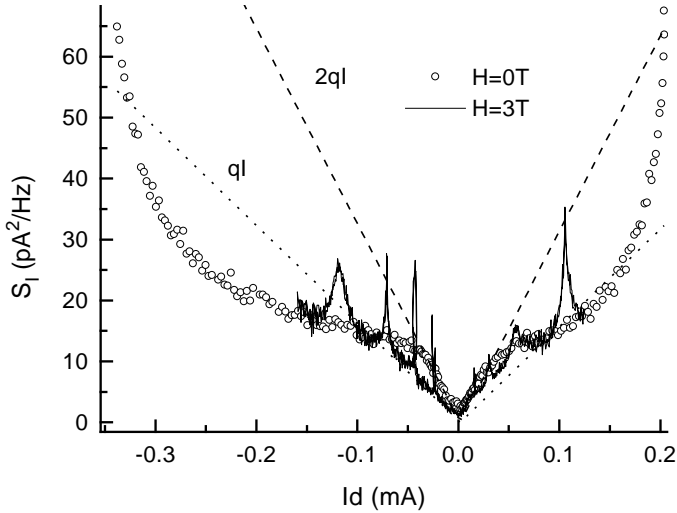}
\vspace{0.2cm}
\caption{The spectral density of the shot noise as a function of current,
measured  at H=0 (circles) and at 3T field (solid line). Two straight lines show
the full shot noise value (dashed line) and half of its value (dotted line). The
peaks in shot noise at 3T correspond to the tunneling through Landau levels in
the well.}\end{figure}

The correlation between peaks in the noise and the features in the I-V
characteristic is apparent in Fig. 3: the voltages at which the noise peaks
occur correspond to those of negative (or close to negative) differential
conductance in the current. 
Moreover, the values of the peak maxima in the noise
spectrum are correlated with the values of concomitant differential conductance
features: the smaller the conductance, the larger the noise peak. Thus, when the
conductance is small but still positive, the noise peak is above the suppressed
noise value $qI$ but below $2qI$. (See, for instance, the peak at about -0.08V.)
When the conductance is around zero full shot noise is approximately recovered.
(See the peak at 0.08V). And when the conductance becomes negative (for example,
at -0.04V and -0.06V) the noise is enhanced well beyond $2qI$.  It should also
be noted that the peaks that show this enhancement pass the $2qI$ value at
voltages at which the conductance is approximately zero.

This behavior can be explained by realizing that noise is sensitive to the
variation of the density of states with a change of the number of electrons in
the well \cite{Iann98}. Let us consider first the case of zero magnetic field,
in which the density of states is constant above the edge provided by the
quantum-state energy.  The tunnel current increases as long as the voltage is
such that the quantum-state energy is below the effective edge of the valence
band of the emitter electrode (Upper inset of Fig. 1).  When a hole from the
GaSb emitter tunnels into the InAs well, the density of states in the well goes
down and the number of states available for subsequent tunneling holes is
smaller.  In this case, there is a correlation between tunneling holes (in other
words, there is a negative feedback between tunneling holes) and noise is
suppressed.  But when the voltage is such that the quantum-state energy is just
below the emitter valence-band edge, the feedback is positive: by going down in
energy when a hole tunnels, the density of states is now increased and
additional holes can tunnel.  This opposite correlation between holes enhances
shot noise, and the rapid increase of noise in Fig. 2 beyond 0.16 mA (for
positive current) and -0.25 mA (for negative current) reflects such an
enhancement.

\begin{figure}
\epsfclipon
\epsfxsize 0.5\textwidth
\epsffile{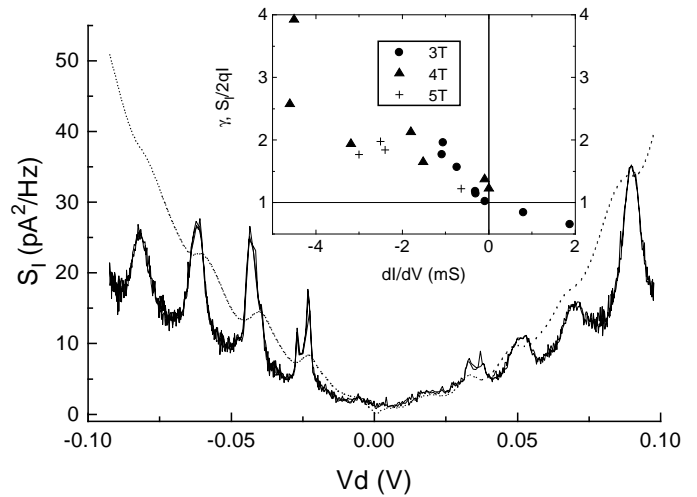}
\vspace{0.2cm}
\caption{Noise spectral density as a function of the voltage across the diode,
at 3T.The dashed line shows calculated full shot noise $2qI$. The insert shows
the ratio of the peak maxima of the shot noise (circles-3T, triangles-4T and
crosses-5T) to $2qI$.}
\end{figure}

An external magnetic field modifies drastically the density of states in the
well from a step function to a series of quasi-delta functions (Landau levels).
Although for the moderate fields considered here it may be more appropriate to
speak of a step function modulated by an oscillatory component, for simplicity
in this discussion we assume a density of states that consists of Gaussian
distributions centered at the Landau-level energies, the widths of which are a
measure of the energy broadening of the states.  With increasing voltage, the
successive alignment of the emitter edge with the centers of those Gaussians
gives rise to peaks in the tunnel current, followed by regions of negative
differential conductance. Each of these regions, then, results from an alignment
of the emitter with a decreasing density of states.  If a hole tunnels now into
the well, the density of state goes down and the emitter becomes aligned with a
portion of the distribution that can accommodate more states, thus favoring the
tunneling of a second hole and therefore enhancing shot noise.  The sharper the
density-of-states distribution (because of higher field or reduced broadening),
the more pronounced is the NDC and the stronger is the enhancement of noise.

The direct connection between shot noise and negative differential conductance
is seen in the insert of Fig. 3, where we have plotted the Fano factor for the
maxima of the noise peaks versus $dI/dV$, for three values of the magnetic
field.  The graph shows a monotonic increase of noise above $2qI$ ($\gamma > 1$)
with increasingly negative conductance, from $\gamma = 1$ when the conductance
is zero.  Although the value of $\gamma$ seems to level off at around 2 before
resuming its increase up to a maximum value of 4, more experiments are needed
before the details of such a behavior are confirmed.  Previous experiments at
zero field have yielded for $\gamma$ values between 2 and 9 (Refs. 2, 7, 8).

Iannacone et al. \cite{Iann98} have expressed the Fano factor in terms of the
characteristic times $\tau_g$, $\tau_r$ for the processes of generation and
recombination of electrons in the well:
\begin{equation}
\gamma\equiv\frac{S_I(\omega)}{2qI}=1-\frac{2\tau_g\tau_r}{(\tau_g+\tau_r)^2},
\label{eq2}
\end{equation}
and used this expression to explain noise enhancement. When $\tau_g$ is zero
full shot noise is recovered, and as it becomes negative in the NDC region
(while $\tau_r$ remains positive) shot noise is enhanced. This expression can be
understood in simple terms with the two-resistance analogy mentioned above, by
which the characteristic times of Eq. (2) are replaced by the differential
resistances of the two barriers. Then, the resistance of the emitter barrier is
infinite at the resonance and negative in the NDC region, and $\gamma$ becomes
one and larger than one, respectively.

In conclusion, we have shown that by changing the shape of the density of states
in the quantum well with a magnetic field the value of the shot noise relative
to the classical value can change from suppression to enhancement. A
quantitative comparison between the prediction of Eq. (2) and the experimental
results of Fig. 3 is not possible without knowing the transition rates as a
function of the number of holes in the well. A calculation of these rates along
the lines of  Ref. 8 is not trivial, especially  in the presence of a magnetic
field. In any case, the validity of Eq. (2) is somewhat surprising since it
follows from a linearization of the transition rates in the NDC region, in which
large charge fluctuations are expected \cite{last}. A microscopic theory that
includes these non-linear effects seems therefore needed.

The authors have benefited from discussions with R. Landauer. This work has been
sponsored by the Department of Energy (DOE's Grant No. DE-FG02-95ER14575).


\begin{references}
\bibitem{notes}Equivalently, the process can be seen as electrons tunneling out
of the InAs well to the valence band of GaSb, followed by electrons from the
other GaSb electrode tunneling into the InAs well.  See, for example, Ref. 12.
\end{references}
\end{document}